# Strong Hot Carrier Effects in Single Nanowire Heterostructures

Iraj Abbasian Shojaei, Samuel Linser, Giriraj Jnawali, N. Wickramasuriya, Howard E. Jackson, Leigh M. Smith
*Department of Physics, University of Cincinnati, Cincinnati, OH 45221*

Fariborz Kargar, Alexander A. Balandin
*Department of Electrical and Computer Engineering, University of California, Riverside CA 92521 USA*

Xiaoming Yuan
*School of Physics and Electronics, Hunan Key Laboratory for Supermicrostructure and Ultrafast Process, Central South University, 932 South Lushan Road, Changsha, Hunan 410083, P. R. China*

Philip Caroff, Hark Hoe Tan, and Chennupati Jagadish
*Department of Electronic Materials Engineering, Research School of Physics and Engineering, The Australian National University, Canberra, ACT 2601, Australia*

## ABSTRACT:

We use transient Rayleigh scattering to study the thermalization of hot photoexcited carriers in single $GaAs_{0.7}Sb_{0.3}$ / InP nanowire heterostructures. By comparing the energy loss rate in single core-only $GaAs_{0.7}Sb_{0.3}$ nanowires which do not show substantial hot carrier effects with the core-shell nanowires, we show that the presence of an InP shell substantially suppresses the longitudinal optical phonon emission rate at low temperatures which then leads to strong hot carrier effects.

Keywords: Hot Carrier Effects, Hot Phonons, Carrier Thermalization

**Introduction:**

The understanding of thermalization processes in semiconductors has significant implications for a wide variety of applications. For instance, one of the barriers to higher efficiency solar cells is the substantial loss of solar energy to heat as carriers thermalize before they are collected electrically.[1–5] When electrons and holes are created with substantial excess energy, they first rapidly relax (< 1 ps) through the emission of longitudinal optical (LO) phonons through polar Frohlich interactions.[6] This creates a large non-thermal distribution of LO phonons which must equilibrate to the lattice temperature. In many materials, such as GaAs, GaSb and the ternary alloy GaAsSb, these non-thermal LO phonons can efficiently thermalize through the Klemens channel where a single zone center LO phonon decays into two counter propagating longitudinal acoustic (LA) phonons. However, in materials such as InP where $\hbar\omega_{LO} > 2\hbar\omega_{LA}$, this thermalization process is inhibited and so the LO phonon populations can remain in a non-thermal state for extended periods which inhibits cooling of the electrons and holes resulting in a pronounced hot carrier effect.[7-9]

The subject of hot carrier effects in semiconductors has been of strong interest for the past twenty years since it was first realized that LO phonon emission could be strongly suppressed in certain materials or structures.[10–12] Substantial progress in understanding hot carrier thermalization in bulk and two dimensional materials has been made, but much less is known about how nanoscale heterostructures might impact thermalization dynamics.[13,14] Numerous measurements have shown that hot carrier effects are more prominent in InP materials than GaAs-based nanowires (NWs),[9,15] while others have shown that hot carrier effects are more dominant in thinner NWs,[16] and terahertz measurements have shown that defects such as stacking faults can enhance hot carrier effects even further.[17–19]

In this paper, we use transient Rayleigh scattering (TRS) to measure directly the average energy per carrier as a function of time in single NWs at 10 and 300 K and extract the energy loss rate for core-only $GaAs_{0.7}Sb_{0.3}$ NWs and $GaAs_{0.7}Sb_{0.3}$ / InP NW core-shell heterostructures. We show that while the core-only $GaAs_{0.7}Sb_{0.3}$ shows the expected optic-phonon dominated rapid thermalization, the growth of an InP shell surrounding $GaAs_{0.7}Sb_{0.3}$ core exhibits extremely strong hot-carrier effects.

**Sample Morphology and Experimental Setup:**

Core-only $GaAs_{0.7}Sb_{0.3}$ and core-shell $GaAs_{0.7}Sb_{0.3}$ / InP NWs were grown via the vapor-liquid-solid (VLS) method using gold catalysts in a metal-organic vapor phase epitaxy (MOVPE) system.[20,21] Figure 1a shows an SEM image of the morphology of the core-shell NWs. Figure 1b shows a TEM image of a cross-section of a core-shell NW which displays the hexagonal inner core with non-polar {110} facets surrounded by outer truncated triangular-shaped InP shell with {112} facets. The diameter of the $GaAs_{0.7}Sb_{0.3}$ core is 70 nm and the thickness of the InP shell is a maximum of 30 nm. Details on the growth and morphology of the core-only and core-shell NWs can be found in Refs. 20 and 21.

Under lattice-matched conditions $GaAs_{0.5}Sb_{0.5}$ / InP heterostructures have been shown to display a type II band alignment (see Figure 1c) with holes confined to the $GaAs_{0.5}Sb_{0.5}$ valence band (VB) with a 0.4 eV InP barrier, while the electrons are confined to the InP conduction band (CB) with a relatively modest $GaAs_{0.5}Sb_{0.5}$ barrier.[22] In $GaAs_{0.7}Sb_{0.3}$ / InP NWs, the $GaAs_{0.7}Sb_{0.3}$ core is under tensile strain while the InP shell is under compressive strain. It is expected that the type I to type II transition should occur at approximately 30% to 40% Sb composition. Thus, our expectation is that the holes are strongly confined to the $GaAs_{0.7}Sb_{0.3}$ core, while the electrons see nearly flat-band conditions in the CB between the core and shell.

In the TRS experiments, a Coherent Chameleon Ti:Sapphire laser with 150 fs 1.5 eV pulses is used to excite single NWs with light polarized parallel to the long axis. The power of the pump beam is kept low enough so that there is no potential for heating the lattice. The change in the polarized scattering efficiency is monitored by a delayed mid-infrared output pulse (150 fs) from a Coherent Chameleon OPO laser with energies ranging from 0.79 to 1.13 eV (1100 to 1570 nm). The polarization of the probe beam oscillates at 100 kHz between parallel and perpendicular to the NW axis using a photoelastic modulator, and the scattered light from the NW is detected using a $LN_2$-cooled InSb detector and a lock-in amplifier. Using a mechanical delay line, the probe pulse can be delayed relative to the pump pulse (-100 to 2000 ps) to investigate the effect of carrier decay and thermalization on scattering efficiency. For this measurement, the pump pulse train is chopped at 800 Hz and the output of the first lock-in is detected by a second lock-in amplifier tuned to this frequency. The second lock-in thus measures the change in the polarized scattering efficiency due to the presence of the photoexcited carriers: $\Delta R' = R'_{on} - R'_{off}$ where $R' = R_{\parallel} - R_{\perp}$. The normalized TRS scattering efficiency, $\Delta R'/R' = (R'_{on} - R'_{off})/R'_{off}$ has a derivative-like behavior as a function of energy and depends on the geometry of the NW and the change of both the real and imaginary parts of complex index of refraction which is function of energy, carrier density and temperature.[9,23,24]

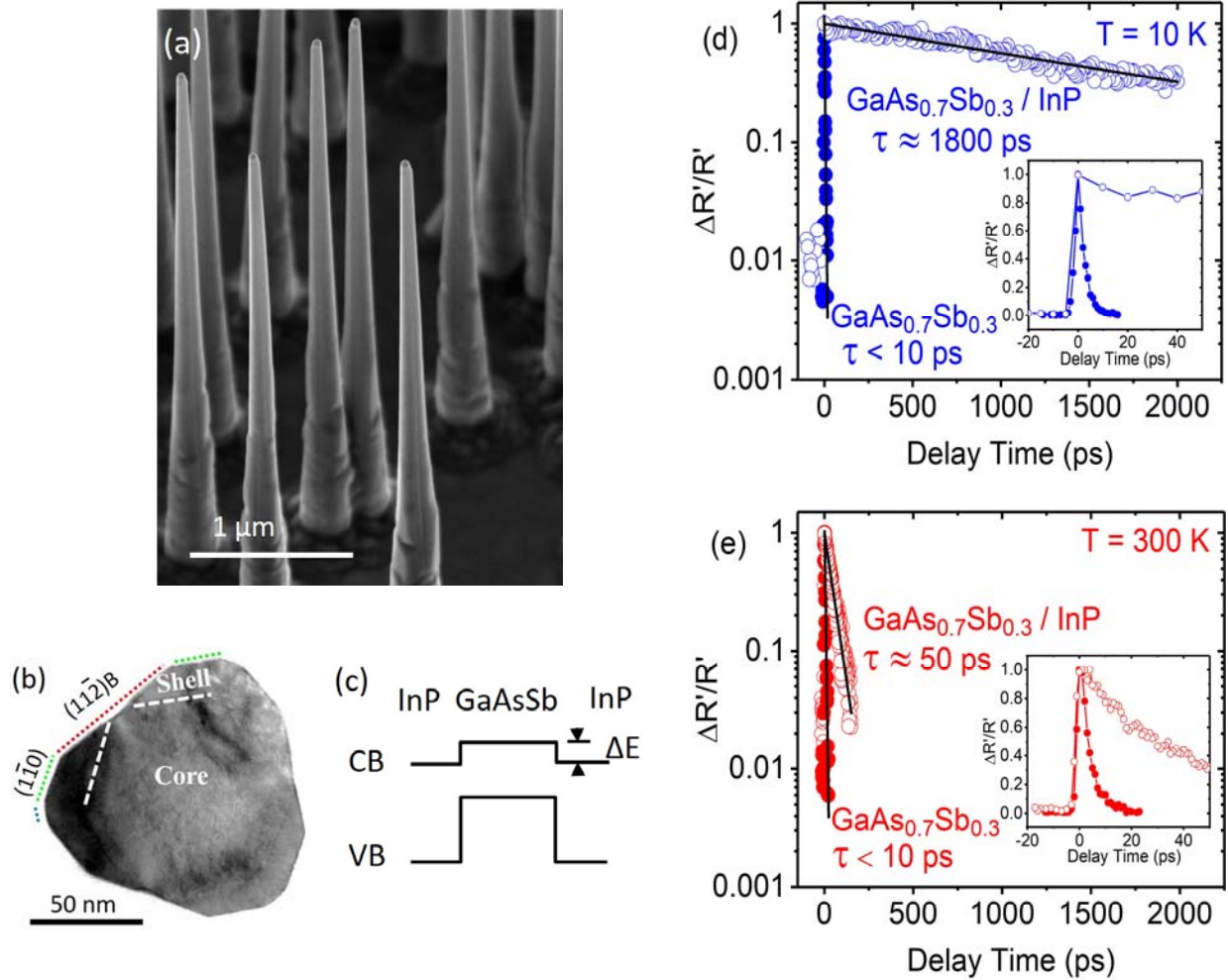

Figure 1: (a) SEM images GaAs$_{0.7}$Sb$_{0.3}$ / InP nanowires. Scale bar is 1 μm. (b) Cross sectional TEM image of GaAs$_{0.7}$Sb$_{0.3}$ / InP NW. Scale bar is 50 nm. (c) Schematic band diagram of type II GaAs$_{1-x}$Sb$_x$ / InP nanowire. (d, e) Solid black lines show exponential decay fitting of selected transient Rayleigh scattering carrier dynamics data for core-only (solid circle) and core-shell (open circle) nanowires at (d) 10 K (blue) and (e) 300 K (red) at fixed 0.83 eV probe energy.

**Experimental Results:**

Figure 1d shows the time decay of the TRS scattering efficiency for the core-only (solid circles) and core-shell (open circles) NWs at 10 K at a fixed 0.83 eV probe energy just below the band edge. The core-only NW decays rapidly to background with a 10 ps exponential decay, while the core-shell NW shows a remarkably long 1800 ps exponential decay. This indicates that photoexcited carrier recombination in the core-only GaAs$_{0.7}$Sb$_{0.3}$ NW is dominated by nonradiative surface recombination. The InP shell clearly passivates the surface states of the GaAsSb core, resulting in a two orders of magnitude longer lifetime. In contrast, at room temperature (300 K), the core-only NW shows a similar 10 ps fast decay, while the core-shell exhibits only a 150 ps time decay (Figure 1e). This is consistent with the CB of the InP shell providing only very weak confinement for electrons of ~30 meV relative to the GaAs$_{0.7}$Sb$_{0.3}$ CB edge. If the band alignment is Type-I with electrons and holes confined to the core, the 30 meV confinement potential in the conduction band would not be sufficient to confine the electrons to the core if they have a thermal energy of ~30 meV at 300 K and so they would see the unpassivated surface states of the InP. If the band alignment is Type-II with the electrons confined to the InP, at low temperatures the Coulomb attraction of the electrons to the strongly confined holes in the core is sufficient to keep them close to the heterointerface. However, at 300 K it is expected that the thermal energy of the electrons would enable them to scatter more frequently with the surface states in the InP. Whether the weak confinement in the CB is type I or II is not possible to determine from our experimental results.

By measuring the energy dependence of the polarized scattering efficiency one can obtain a more detailed understanding of the carrier dynamics.[25,26] As noted previously, the resulting line-shapes exhibit a derivative-like spectra where the zero-crossing and linewidths depend directly on the density and temperature of the electron hole pairs, and the diameter of the nanowires. Using the analysis described in refs (23, 24) and detailed in the Supplemental Information, we are able to extract directly the density and temperature of the electron-hole pairs as a function of time after photoexcitation by the pump pulse.

Figure 2 a,b shows such TRS spectra measured at ~10 K at three different delay times of the probe pulse for (a) GaAs$_{0.7}$Sb$_{0.3}$ core-only and (b) GaAs$_{0.7}$Sb$_{0.3}$ / InP core-shell NWs. As described by Montazeri et al.,[23,24] the zero crossing point of NWs TRS spectra occurs approximately at the band gap of the structure. The line-shape fit is sensitively dependent on the diameter of the nanowire, and the density and temperature of the electron and hole distributions. The core-only NWs display a zero crossing at ~ 0.9 eV which shifts slightly to lower energy at later times. The line-shape for the core-only NW is very broad at early times and narrows within 20 ps which is indicative of filling of the conduction and valence bands with a dense and hot electron hole plasma which decays and thermalizes rapidly. The behavior of the core-shell NW is dramatically different with a smaller line-shape narrowing which extends over a two-orders of magnitude longer time of 2000 ps.

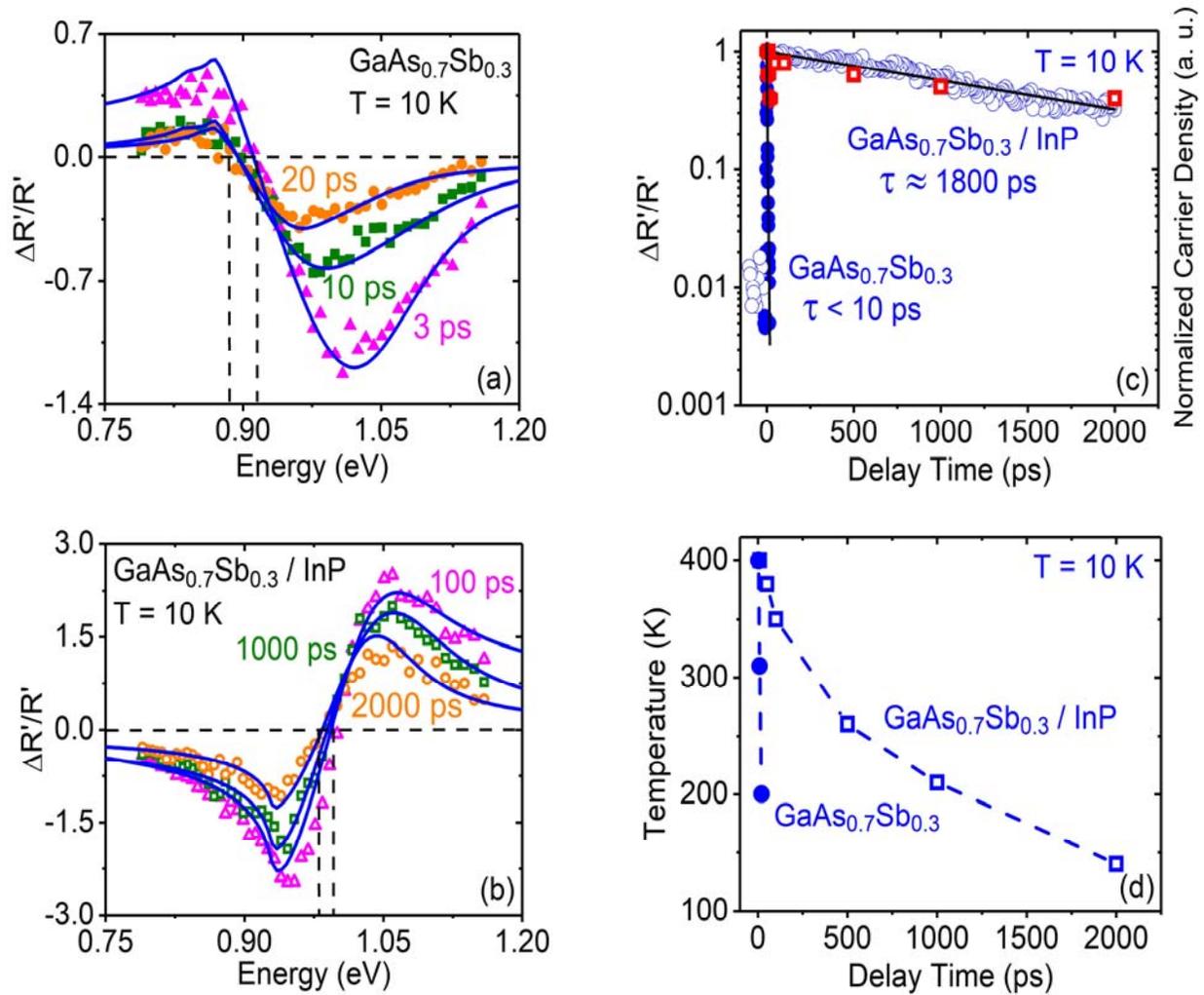

Figure 2: Theoretical fitting (blue lines) of transient Rayleigh scattering spectra at three different time delays at 10 K for (a) core-only (70 nm diameter) and (b) core-shell (130 nm diameter) nanowires. (c) Normalized carrier density of core-only and core-shell nanowires at 10 K have been marked on transient Rayleigh scattering time scan data with red hexagons and red squares respectively. Initial carrier density for core-only nanowires is around $4 \times 10^{18}$ cm$^{-3}$ and for core-shell nanowires is about $2 \times 10^{18}$ cm$^{-3}$. (d) Carrier temperature from modeling of transient Rayleigh scattering spectroscopy data at different measured delay time for core-only and core-shell nanowires at 10 K.

By minimizing the difference between the theoretical line-shape and the data points (see section S1 in Supplemental Information), we are able to determine the electron-hole density and temperature for each delay time. The diameter of the nanowire is determined so that the chi-square is minimized for all spectra. In the fits displayed in Figure 2 the NW diameters determined in this way are 70 and 130 nm for core-only and core-shell NWs respectively, which are consistent with cross-sectional TEM measurements.[21] Using these NW diameters and assuming the number of light and heavy holes equals the number of electrons, it is straightforward to determine the carrier density and temperature from each spectrum (see solid lines in Fig. 2 a and b). The time dependent density and temperature extracted from the fits of spectra at 10 K are shown in Figure 2c,d. The red squares (red hexagons) in Figure 2c show the normalized carrier densities in the core-shell (core-only) NWs at times after photoexcitation which is in good agreement with the time decays shown previously. The initial carrier density for core-only NWs is ~ $4\times10^{18}$ cm$^{-3}$ and ~ $2 \times 10^{18}$ cm$^{-3}$ for the core-shell NWs. The fits confirm that the carrier density in the core-shell NWs takes 600 times longer to decay. This suggests that the band alignment of the core and shell is type-II (with holes confined to the core), but is not conclusive. The temperature (Figure 2d) of the photoexcited electrons and holes in the core-only NWs drops from 400 to 200 K in 20 ps, while in the core-shell NWs it takes nearly 2000 ps to drop to 140 K from the same initial temperature. These spectra also confirm that the lattice temperature does not change from the nominal 10 K.

Figure 3a,b shows TRS spectra for core-only and core-shell NWs at ~300 K. The behavior of the line-shapes for the core-only NWs is very similar to that at low temperature, while the line-shape for the core-shell NWs exhibits a somewhat faster decay and thermalization with time, consistent with the shorter time decays observed at room temperature. The fits to the 300 K spectra (see Figure 3) again result in the same 70 and 130 nm diameters for the core-only and core-shell NWs respectively. The density and temperature extracted from these fittings at room temperature are shown in Figure 3c,d respectively. The normalized carrier densities obtained from the fits are shown in Figure 3c with blue hexagons for core-only NWs and blue squares for core-shell NWs. These points display good agreement with the related TRS time scan. The initial densities are $5\times10^{18}$ cm$^{-3}$ for the core-only NWs and $6.3\times10^{18}$ cm$^{-3}$ for the core-shell NWs. These fits show extremely rapid carrier thermalization in the core-only NWs from 500 K to 320 K within ~20 ps after the pump pulse. The core-shell NWs, on the other hand, show a slower thermalization from 550 to 320 K within 150 ps (Figure 3d).

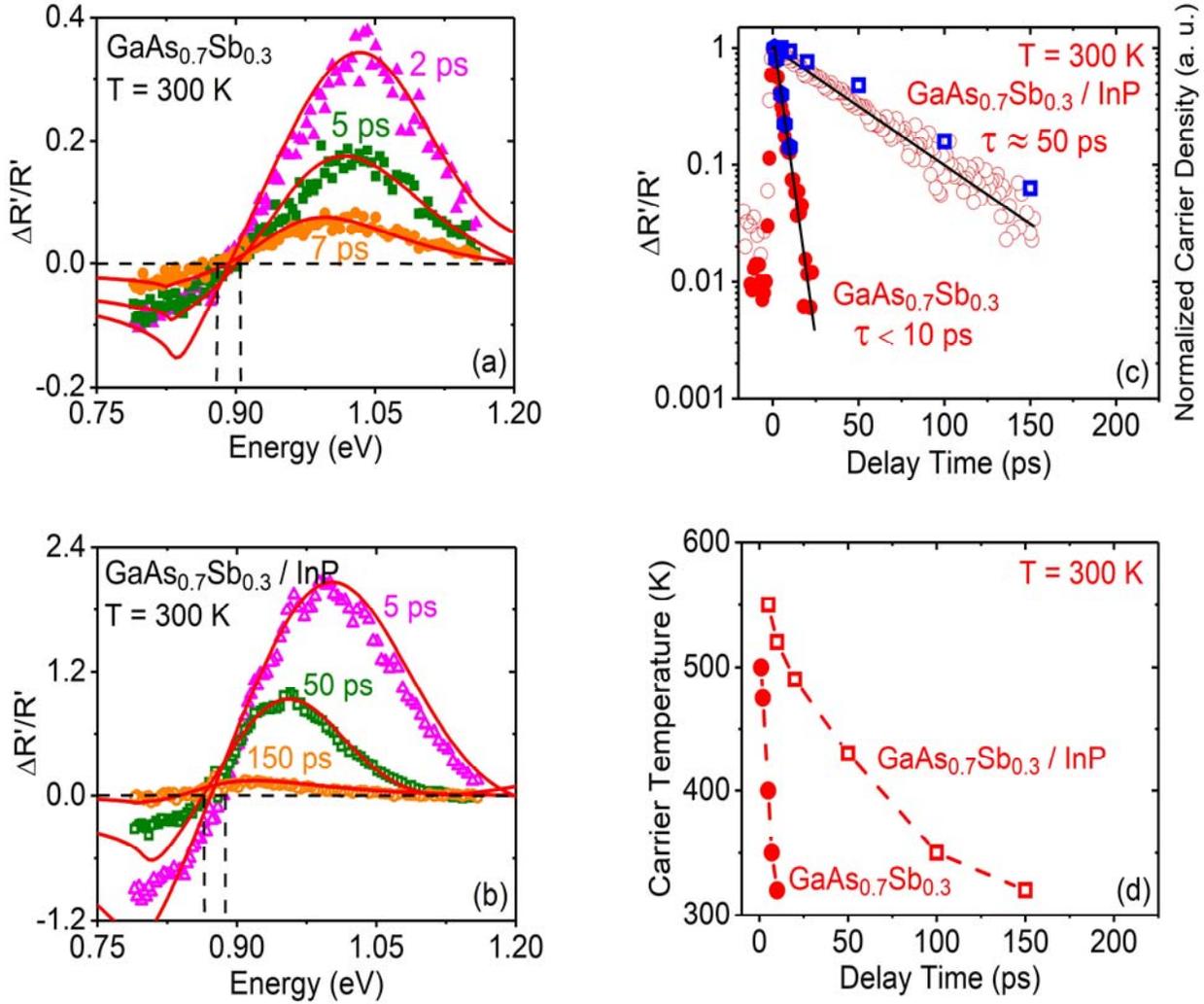

Figure 3: Theoretical fitting (red lines) of transient Rayleigh scattering spectra at three different time delays at 300 K for (a) core-only and (b) core-shell nanowires. (c) Normalized carrier density of core-only and core-shell nanowires at 300 K have been marked on transient Rayleigh scattering time scan data with blue hexagons and blue squares respectively. Initial carrier density for core-only nanowires is around $5 \times 10^{18}$ cm$^{-3}$ and for core-shell nanowires is about $6.3 \times 10^{18}$ cm$^{-3}$. (d) Carrier temperature from modeling of transient Rayleigh scattering spectroscopy data at different measured delay time for core-only and core-shell nanowires at 300 K.

From the fits to the time-resolved scattering spectra of these NWs suggest several conclusions. The first is that the InP shell clearly passivates non-radiative surface states in the GaAs$_{0.7}$Sb$_{0.3}$ NWs at both 10 and 300 K resulting in substantially longer recombination lifetimes in the core-shell NWs. The much larger lifetime enhancement observed at low temperatures may indicate that the band alignment of the core-shell NW is marginally type-II with electrons confined to the InP with a 30 meV confinement energy. However, the carrier temperature dynamics also indicate that the presence of the InP shell clearly causes a substantial slowing of the thermalization times although GaAsSb (like GaAs) is not known to be a material which shows substantial hot carrier effects. In the following sections we quantify the change in the energy loss rate in the core-shell NWs.

**Carrier Thermalization:**

Through the fitting process described above we can determine the electron and hole densities (their quasi-Fermi energies) and temperature as a function of time. We can therefore calculate the dynamic change in the average energy per carriers using the expression:[27–29]

$$E = \frac{3}{2} k_B T \frac{F_{\frac{3}{2}}(\eta)}{F_{\frac{1}{2}}(\eta)} \quad (1)$$

where $\eta$ is the quasi-fermi energy and $F_i(\eta)$ is the $i^{th}$ Fermi integral defined in the usual manner. Using this result and the measured dynamics of the temperature and Fermi energies for electrons and holes we can calculate the average energy per electron-hole pair as a function of time after photoexcitation for both the core-only and core-shell NWs.

The thermalization of electrons and holes is determined by the scattering (emission) rate of the carriers with LO and LA phonons which determines their energy loss rate.[11,30,31] From the average energy per pair, we can calculate the energy loss rate (ELR) versus time simply by calculating the numerical derivative. The ELR calculated in this way is shown in Figure 4a,c for the core-only NWs at 10 K and 300 K respectively and Figure 4b,d for the core-shell NWs at 10 K and 300 K respectively. The difference between the core-only and core-shell NWs at 10 K is immediately obvious as the ELR of the core-only is three orders of magnitude larger than that of the core-shell NWs.

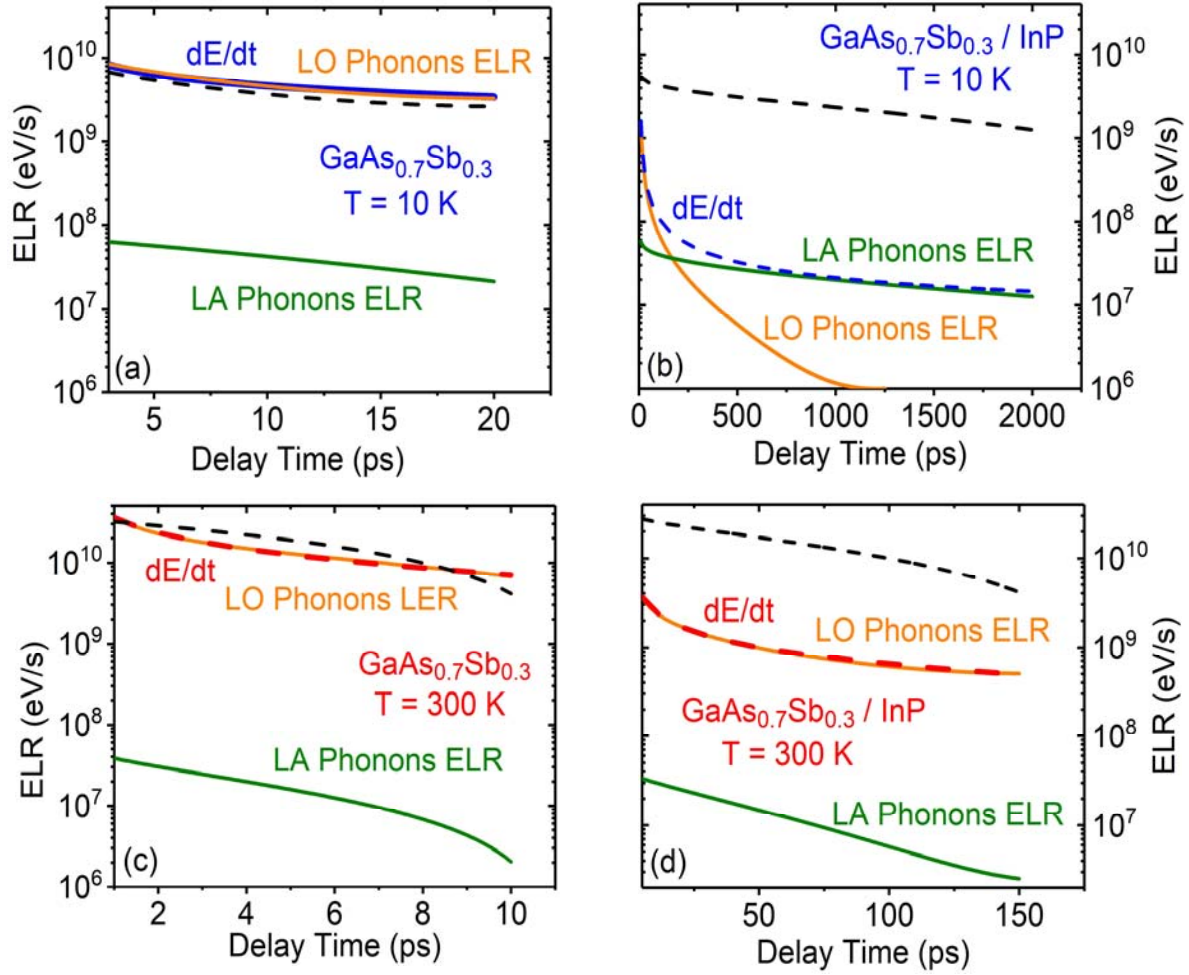

Figure 4: dE/dt and carrier energy loss rate due to optical and acoustic phonon emission for core-only and core-shell nanowires respectively at 10 K (a, b) and 300 K (c, d). Dashed lines in these graphs show energy loss rate due to optical phonon emission calculated by the Ridley expression.

The reduction in the energy per particle reflects the thermalization of the carriers as a function of time. The derivative dE/dt also shows the dynamics of the ELR as a function of time. Because the thermalization process is dominated by emission of both LO and LA phonons, it is clear that:[28–30]

$$\langle \frac{dE}{dt} \rangle = c \ \langle \frac{dE(N(t),T(t))}{dt} \rangle_{LO} + \ \langle \frac{dE(E_{ac},N(t),T(t))}{dt} \rangle_{LA} \quad (2)$$

The left hand side of the equation is determined directly from the TRS measurements. Hot carrier effects in semiconductors occur because of a large suppression of the LO phonon emission rate because of hot phonons which cannot down convert efficiently to LA phonons.[27,28] On the other hand, if one knows the acoustic deformation potential, the ELR for LA phonons is quite well understood.[32] This means that it is possible to extract the LO phonon ELR simply by subtracting the ELR for LA phonons directly from dE/dt calculated from the TRS data.

For example, dE/dt from the TRS data for the core-only NWs at 10 K shows a total ELR which is ~$10^{10}$ eV/s. Given the deformation potential of 1.6 eV, the LA phonon ELR is just below $10^8$ eV/s.[9] This means that the thermalization of hot carriers in the core-only NWs is completely dominated by LO phonon emission which explains the rapid decrease in temperature of the carriers (see Figure 4a). In contrast, the 10 K measurements for the core-shell NWs show a radically different behavior. While the total ELR starts at $10^9$ eV/s, it falls rapidly to mid $10^7$ eV/s and decreases slowly after that. We adjust the deformation potential to 1.6 eV in order to fit the late time response of the energy loss rate for the 10 K core-shell nanowires. By subtracting the LA phonon ELR from dE/dt, we therefore obtain the dynamics of the change in the LO phonon ELR in the core-shell NWs. This shows that the LO phonon ELR dominates at times less than 200 ps after the pump pulse but falls rapidly below the LA phonon ELR at later times (see Figure 4b).

The black dashed lines in Figure 4 display the ELR due to LO phonon emission based on Ridley expression.[27,33–35] In this calculation, the LO phonon ELR depends on carrier temperature and density, lattice temperature and reabsorption of LO phonons which is represented by a coefficient in that formalism. For a given 0.0025 reabsorption coefficient, the ELR due to LO phonon emission is close to the carrier ELR in the core-only NW with an excellent correspondence to the LO phonon emission extracted from the dynamic measurements. But in the core-shell NWs by using same value of the reabsorption coefficient (0.0025), the LO ELR is close to our experimental result at early times, but is orders of magnitude too high at times greater than 200 ps.

The LO phonon ELR can be related to the emission rate of the LO phonons through $\hbar\omega/\tau^*$, where $\tau^*$ is the time between LO phonon emissions.[36] This shows that the LO phonons emission rate for the core-only NWs is ~2 ps, while that for the core-shell NWs at 10 K is 10 ps at the earliest times but rapidly increases by three orders of magnitude to 4000 ps by 1 ns after the pump pulse. This result is shown in more detail in Fig. S1 in the Supplemental Information.

Similar analysis of TRS data taken at room temperature (~300 K) shows that the LO phonon emission time for core-only NWs drops to 100 fs, while that for the core-shell NWs remains stable at 5 ps. This implies that even at room temperature hot carrier effects are not negligible in $GaAs_{0.7}Sb_{0.3}$ / InP nanostructures (see Figure 4c,d). The black dashed lines in this graph show the ELR based on the Ridley expression with a 0.025 reabsorption coefficient. This coefficient is 10 times larger than the 10 K value but matches well with the time-dependent ELR extracted from core-only NWs. Comparing the results in Fig. 4 c and d we see that the calculated LO phonon ELR from the core-only NW is almost one order of magnitude larger than the ELR extracted from TRS measurements in the core-shell NW.

**Discussion:**

The central conclusion from the above analysis is that a $GaAs_{0.7}Sb_{0.3}$ semiconductor NW made of material which should *not* show hot carrier effects, now shows very strong hot carrier effects at low temperature if a thin 30 nm InP shell is added to the NW.  In the discussion above, we considered mainly ELR of the charged carriers. We now consider how the phonon populations are affected after excitation by the pump pulse. Electrons and holes are created with nearly 600 meV of excess energy. This means that nearly 20 hot phonons per pair are created by the rapid relaxation of hot electrons and holes to the band edge through the Frohlich interaction. This is potentially different in the core-shell NW because Froehlich coupling in the InP shell is three times larger than the $GaAs_{0.7}Sb_{0.3}$ core (0.15 vs 0.05).[37] This may suggest that during the initial relaxation of the hot carriers, substantially more InP-like LO phonons are created than $GaAs_{0.7}Sb_{0.3}$ phonons. Secondly, to thermalize with the lattice, these hot phonons need to decay anharmonically to the lower frequency LA phonon branches. Hot carrier effects happen because such anharmonic decays are inhibited and so hot phonons persist for much longer times which in turn inhibits thermalization of the hot carriers.

Over the past decade there has been intense interest in *phonon engineering* whereby one can use nanoscale heterostructures to tune the phonons in a material and also their interactions.[38,39] Spatial confinement of phonons in nanostructures have been shown to strongly impact their dispersion, group velocity and density of states.[38,40,41] While there have been a few papers which claim theoretically that the electron-phonon coupling can be impacted by the presence of a heterostructure, most conclusions are that such an effect is small.[41] Several papers, however, indicate that a shell can strongly impact the LA phonons in the NW, particularly if there is a large impedance mismatch between the core and the shell.[42-46] The impedance, defined as $\eta = \rho v_s$, where $\rho$ is the density and $v_s$ is the sound velocity in the material, is substantially larger (40%) for the $GaAs_{0.7}Sb_{0.3}$ core than for the InP shell. Thus, the outer shell is "softer" than the core, and this has been shown to deplete the density of states of the phonons in the core and confining the phonons to the shell, particularly for large wave-vectors (high frequencies).[42,43] Several papers have shown theoretically that the thermal conduction in both two- and one-dimensional structures can be strongly suppressed with the addition of a softer cladding layer.[44,45] Others have shown that the mobility in the core can be enhanced by suppression of the LA phonon modes in the core.[41] Similarly, Stroscio and Dutta have shown that in a nanostructure where the LO phonons are confined (keeping LA phonons not confined) that the anharmonic decay of the LO phonons is suppressed, resulting in a factor of two longer lifetimes.[47]

The question is whether in the present case of a 70 nm diameter $GaAs_{0.7}Sb_{0.3}$ core and a 30 nm thick InP shell such confinement effects could be relevant. Two estimates of size scales where phonon confinement effects can be seen are when the diameter of the NW is either comparable to $\lambda_T = h v_s / k_B T$ or the diameter of the NW is less than the phonon mean free path, $d_{MFP} = 3K / (C_V v_s)$, where K is the thermal conductivity, $C_v$ is the specific heat and $v_s$ is the phonon velocity. We use tabulated values for these parameters.  At 10 K the thermal wavelength is approximately 20 nm while the mean free path is typically >200 nm. Recently Balandin and co-workers have shown using Brillouin scattering spectroscopy that confinement of LA phonons can be observed at room temperature in GaAs NWs with diameters as large as 130 nm.[48]  It has also been demonstrated experimentally that LA phonon confinement may affect thermal transport in nanostructures with the feature sizes of 25 nm.[49] While the exact mechanism of the phonon dispersion modification on the charge carrier relaxation requires a separate theoretical investigation, one can conclude that confinement of phonons is certainly applicable for the 30 nm InP shell, and the core can produce some effects at low temperatures. The fact that the hot carrier effects in the core-shell NWs are stronger at low temperature may reflect the temperature dependence of the thermal phonon wavelength and the mean free path.

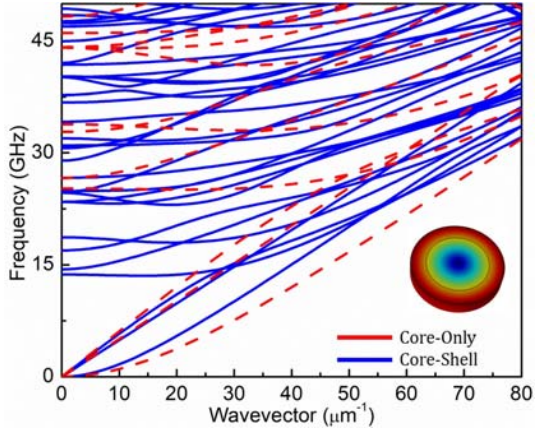

Figure 5: Calculated phonon dispersion in nanowires with and without a shell layer. The red and blue curves show the phonon dispersion along the nanowire axis in a GaAs$_{0.7}$Sb$_{0.3}$ nanowire with D=70 nm and GaAs$_{0.7}$Sb$_{0.3}$ / InP nanowire with the same inner diameter with an InP shell layer of 30 nm thickness.

In Figure 5, we present the results of numerical simulations of the LA phonon dispersion in these NWs with a core diameter of 70 nm and a shell thickness of 30 nm. The simulations reveal noticeable differences between core-only and core-shell NWs, which potentially might be responsible for the experimentally observed phenomena. We see that the shell causes a significant change in the dispersion of all phonon polarization branches. In particular, one can see a decrease in the group velocity of the LA phonon polarization branch. The shell layer also induces bending of the LA phonon branch at smaller phonon wave-vectors compared to the NW without shell layers. The latter translates into substantially lower LA phonon energy at the Brillouin zone edge. While the differences among the phonon branches in terms of the absolute values of energy may not be large, the changed phonon group velocity and density of states can produce measurable effects at low temperature. This can be understood from the following considerations. The electron – phonon scattering via deformation potential depends on the divergence of the phonon displacement (see for example Refs. [44-45]). The phonon displacement in a nanowire is different from that in bulk crystals and would depend on the diameter of the nanowire and mismatch between the nanowire core and the shell. This suggest that the electron relaxation will have a functional dependence on the specifics of the phonon dispersion, particularly at low temperature.

**Conclusions:**

We have shown using TRS to probe core-only GaAs$_{0.7}$Sb$_{0.3}$ and GaAs$_{0.7}$Sb$_{0.3}$ / InP core-shell NWs that the presence of the InP shell *strongly* influences hot carrier effects in these structures. For the core-only GaAs$_{0.7}$Sb$_{0.3}$ no hot carrier effects are seen, and the thermalization of photoexcited carriers is completely dominated by optical phonon emission at both 10 and 300 K. On the other hand, in the GaAs$_{0.7}$Sb$_{0.3}$ / InP core-shell NW at 10 K the LO phonon emission is completely suppressed at times longer than 200 ps and so thermalization is determined almost completely by LA phonon deformation potential scattering. At 300 K, thermalization of hot carriers in the core shell NW is determined by the LO phonon emission, but strong hot carrier effects are still observed with the emission rate reduced by an order of magnitude from the core-only NW. This provides the first evidence that it might be possible to use concepts from *phononic engineering* to control hot carrier effects in semiconductors.

Supporting Information. Description of fitting procedure of transient Rayleigh scattering lineshapes used to extract the radius of the nanowire and density and temperature of the photoexcited carriers. Graphic which shows optic phonon lifetimes versus time after pump pulse.

**ACKNOWLEDGEMENTS:**

We acknowledge the financial support of the NSF through grants DMR 1507844, DMR 1531373 and ECCS 1509706, and also the financial support of the Australian Research Council. The Australian National Fabrication Facility is acknowledged for access to the growth facility used in this work. A.A.B. acknowledges the support of DARPA project W911NF18-1-0041. X.M. Yuan thanks the financial support of the National Natural Science Foundation of China (No. 51702368).

# SUPPLEMENTAL INFORMATION:

# Strong Hot Carrier Effects in Single Nanowire Heterostructures

Iraj Abbasian Shojaei, Samuel Linser, Giriraj Jnawali, N. Wickramasuriya, Howard E. Jackson, Leigh M. Smith
*Department of Physics, University of Cincinnati, Cincinnati, OH 45221*

Fariborz Kargar, Alexander A. Balandin
*Department of Electrical and Computer Engineering, University of California, Riverside CA 92521 USA*

Xiaoming Yuan
*School of Physics and Electronics, Hunan Key Laboratory for Supermicrostructure and Ultrafast Process, Central South University, 932 South Lushan Road, Changsha, Hunan 410083, P. R. China*

Philip Caroff, Hark Hoe Tan, and Chennupati Jagadish
*Department of Electronic Materials Engineering, Research School of Physics and Engineering, The Australian National University, Canberra, ACT 2601, Australia*

## S1. Theoretical Fitting:

As discussed in M. Montazeri et al and Y. Wang et al,[1-3] the line-shape of TRS spectra is sensitive to the three parameters: the density and temperature of the electron-hole plasma and the diameter of the NW. Since the NW diameter is much smaller than wavelength of the probe beam, back scattered light from NW is in the Rayleigh scattering regime. Also, because of large ratio of length to diameter of the NW and dielectric contrast of the NW with surroundings (air), polarization dependent classical Rayleigh scattering from a long uniform cylinder of radius r is a reliable approximation for analysis of scattered light from the NW. Using the Maxwell equations, It has been shown that scattered light by an infinite cylinder with radius r for light with incident polarization parallel ($R_\parallel$) and perpendicular ($R_\perp$) to the axis of cylinder depends on cylinder radius r and complex index of refraction (**n**)[4]. In general $R_\parallel \neq R_\perp$ and thus the reflectance of the NW is polarization dependent. Thus, measuring $R´ = R_\parallel - R_\perp$ enable us to distinguish the scattered light of the NW from the large background reflection from the substrate. Since the complex index of refraction varies by with the carrier density and temperature, the carriers photoexcited by the pump beam changes $R´$, and $\Delta R´ = R´_{on} - R´_{off}$ shows the effect of the excited carries on the polarized scattering efficiency of the NW. Through band filling, the absorption (the complex part of the index of refraction) of the NW is modified by the presence of the electrons and holes. In turn, the real part of the index of the refraction is also modified. Thus, the normalized scattering efficiency $\Delta R´/ R´$ at different times after the pump excitation allows us to extract the carrier density and temperature and the diameter of the

NW. The solid blue lines (red lines) in Figure 2a,b (Figure 3a,b) in the paper illustrate the line-shape of the theoretical modeling of $\Delta R'/R'$ based on the following assumptions:

(1) The electrons and holes are in thermal equilibrium at all times because of the extremely rapid carrier-carrier scattering rate.
(2) The sum of the heavy and light hole densities in the VB equals the density of electrons in the CB.
(3) When the pump is off, the background carrier density and temperature is assumed to be $10^{15}$ cm$^3$ at 10 and 300 K.
(4) All spectral fits at all times use the same value for the nanowire radius. (This radius is chosen to minimize the error for all fits).

We calculate the energy dependence of the absorption by using direct band-to-band transition theory when the CB and VB are occupied by hot electrons and holes respectively. As shown previously,[1-3] the TRS efficiency ($\Delta R'/R'$) depends only on the NW diameter and the change in the real and imaginary part of complex index of refraction:

$$\frac{\Delta R'}{R'} = \frac{\Delta(R_\parallel - R_\perp)}{R_\parallel - R_\perp} \sim \text{Re}\,[e^{i\theta}\mathbf{\Delta n}] \quad (1)$$

where $\theta$ is the modulation phase factor that depends on the NW diameter[2], and $\mathbf{\Delta n} = \Delta n + i\,\Delta k$ is the change of the complex index of refraction due to the occupied CB and VB, where n is the index of refraction and k is proportional to absorption, $k = (\lambda/4\pi)\alpha$. Thus the derivative-like line-shape of $\Delta R'/R'$ can be expressed as:

$$\frac{\Delta R'(E,t)}{R'(E,t)} = \text{Re}\left[A\,e^{i\theta}\left(\Delta n + i\frac{\lambda}{4\pi}\Delta\alpha\right)\right] = A\{\cos(\theta(r))[n(E, N_{eh}(t), T_{eh}(t)) - n(E, N_0, T_0)] - \frac{\lambda}{4\pi}\sin(\theta(r))[\alpha(E, N_{eh}(t), T_{eh}(t)) - \alpha(E, N_0, T_0)]\} \quad (2)$$

where A is an overall arbitrary amplitude factor. The individual time-resolved TRS spectra are modeled using this formula to extract the time-dependent carrier density and temperature in addition to the NW diameter. The absorption coefficient $\alpha(E,N,T)$ is calculated using

$$\alpha(E, N, T) = \frac{\pi^2 c^2 h^3}{n^2 E^2 (2\pi)^3} B \int_0^{E-E_g} \rho_c(E')\,\rho_v(E'-E)[f_l(E - E_g - E') - f_u(E')]\,dE' \quad (3)$$

where B is the radiative bimolecular coefficient, n is the average index of refraction where we have used the average values from literature, $\rho_i(E)$ is the 3D density of states in the

CB and VB, $f(E) = (1 + \exp[(E - E_F) / k_B T])^{-1}$ is the Fermi-Dirac distribution probability that upper and lower states involved in the transition are occupied by electrons, with the quasi-Fermi energy $E_F(N,T)$ related to both the carrier density N and temperature T, and E´ is the upper state energy above the CB minimum. Using the Kramers-Kronig relation we transform the calculated absorption coefficient to acquire the index of refraction $n(E,N,T)$ as a function of energy, carrier density and temperature.

To fit the TRS spectra as a function of time, the absorption coefficient and index of refraction ($\alpha$ and n) are calculated as a function of energy, carrier density and temperature (E, N, T) until the spectra are best fit to the resulting line-shapes using the expression for $\Delta R´/R´$. By minimizing the difference between the theoretical line-shape and the data points, we are able to determine the electron-hole density and temperature for each delay time.

## S2. Effective relaxation time of LO phonons:

The effective relaxation time of the LO phonons ($\tau^*$) for core-only and core-shell NWs at both 10 and 300 K extracted from our calculation for ELR (see Fig. 4 in the manuscript) is shown in Figure S1. For core-only NW at 10 K, it exhibit a constant value of 2 ps for the whole relaxation (Figure S1 a). On the other hand, by adding InP shell the $\tau^*$ display a huge change. It starts around 10 ps and after 1 ns we see the value of around 4000 ps (Figure S1 b). At room temperature, also, the InP shell cause that $\tau^*$ increases around one order of magnitude for all relaxation time (Figure S1 c, d). The black dashed lines in Figure S1 shows $\tau^*$ based on Ridley expression which have nice corresponding for core-only NWs, but it is off from our calculations for core-shell NWs.

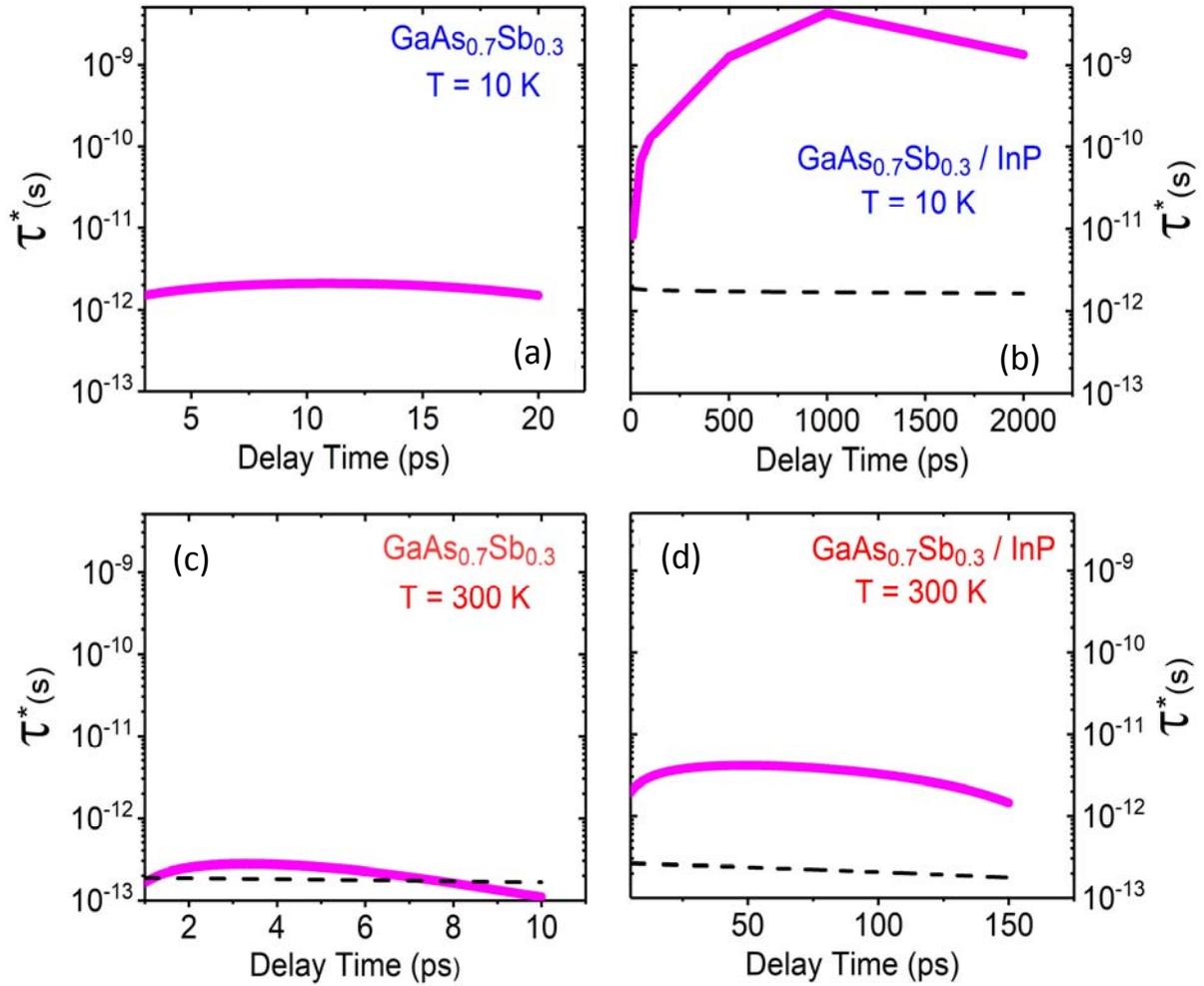

**Figure S1:** The effective relaxation time of optical phonons for the core-only and core-shell nanowires at 10 and 300 K. Black dashed lines in the graph show effective relaxation time of optical phonons for core-shell nanowires calculated by Ridley expression.